\documentclass{ECOC-author}

\usepackage{float}
\usepackage{graphicx, subcaption, caption}

\begin{document}

\title{MODE-MULTIPLEXED FULL-FIELD RECONSTRUCTION USING DIRECT DETECTION AND PHASE RETRIEVAL
\vspace{8pt}}

\author{Haoshuo~Chen\ad{1}, 
        Juan~Carlos~Alvarado-Zacarias\ad{1,2},
        Hanzi~Huang\ad{1,3},
        Nicolas~K.~Fontaine\ad{1},
        Roland~Ryf\ad{1},
        David~T.~Neilson\ad{1},
        and Rodrigo~Amezcua-Correa\ad{2}
}

\address{\add{1}{Nokia Bell Labs, 791 Holmdel Rd., Holmdel, NJ 07733, USA}
\add{2}{CREOL, The Univ. of Central Florida, Orlando, Florida 32816, USA}
\add{3}{Key laboratory of Specialty Fiber Optics and Optical Access Networks, Shanghai Univ., 200444 Shanghai, China}
\email{haoshuo.chen@nokia-bell-labs.com}}

\keywords{MODE-DIVISION MULTIPLEXING, DIRECT DETECTION, PHASE RETRIEVAL}

\begin{abstract}
We realize mode-multiplexed full-field reconstruction over six spatial and polarization modes after 30-km multimode fiber transmission using intensity-only measurements without any optical carrier or local oscillator at the receiver or transmitter.
The receiver's capabilities to cope with modal dispersion and mode dependent loss are experimentally demonstrated.
\end{abstract}

\maketitle

\section{Introduction}
Detecting the full field of all the spatial and polarization modes supported in a multimode fiber (MMF) enables multiple-input multiple-output (MIMO) processing to undo mode coupling~\cite{SDM} and to compensate chromatic dispersion (CD)~\cite{CD} and other transmission impairments.
Most space division multiplexing (SDM) transmission experiments uses coherent receivers~\cite{COH1,COH3} to capture the full electrical field of a signal waveform by measuring its interference against a stable continuous-wave (CW) local oscillator (LO). 
In order to simplify the receiver architecture using only direct detection, the Kramers–Kronig~\cite{KK3,KK10} and Stokes space~\cite{Stokes} receiver schemes have been adapted to support mode-multiplexed transmission.
However, in both schemes, a CW tone is needed somewhere in the system.
For the Kramers–Kronig scheme~\cite{KK3,KK10}, the CW tone is added at the receiver to avoid carrier power fading~\cite{KK_Fading}. 
The Stokes space receiver~\cite{Stokes} cannot fully utilize the mode dimension for transmitting the signals since at least one spatial and polarization mode needs to transmit the CW carrier.

\begin{figure}[!t]
	\vskip -3pt
	\centerline{\includegraphics[width=2.6in]{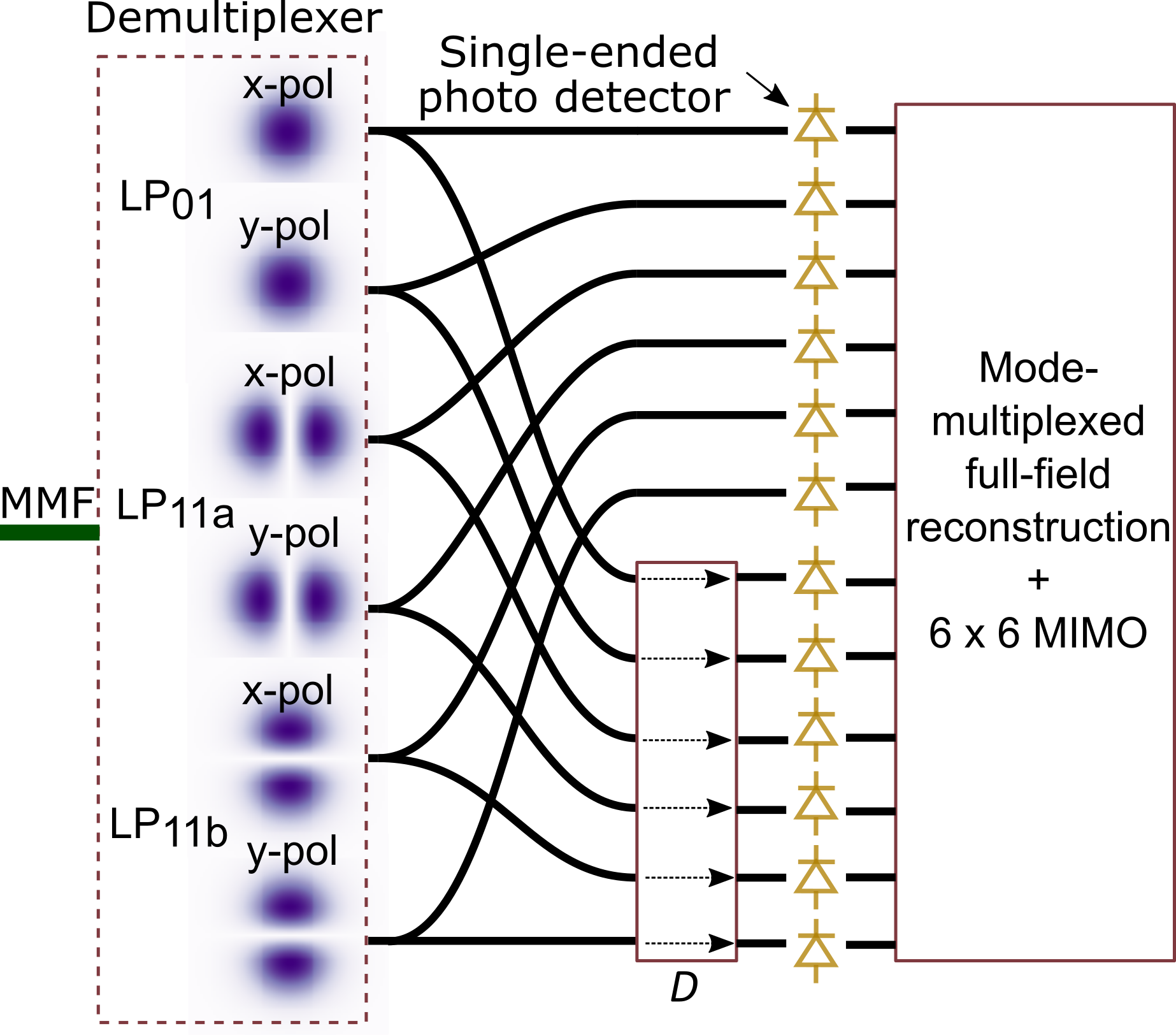}}
	\vspace{-3pt}
	\caption{~Schematic of the direct detection receiver for reconstructing the full field of six spatial and polarization modes using alternative projection for phase retrieval.
	(Dispersion $D$ is applied as the pre-determined alternative projection function).   
	}
	\vspace{-8pt}
	\label{FIG:1}
\end{figure}

The use of CW carrier or LO can be eliminated employing phase retrieval~\cite{PR1,PR2}.
Compressed sensing techniques exploiting signal sparsity are powerful tools for phase retrieval~\cite{PR1}.
However, to produce sparsity from a non-sparse telecommunication signal requires numerous (>>2) linear projections of the signal for each launched spatial and polarization mode, which increases the receiver complexity since each projection requires measurement on a separate receiver~\cite{PR1}.
The receiver size would be extremely large if more parallel signals such as mode-multiplexed signals need to be detected.

In this paper, we propose and demonstrate a scalable phase retrieval receiver capable of reconstructing the full field of mode-multiplexed complex-valued signals.
It is based on our recent polarization-diversity demonstration~\cite{PR2} with improvements such as more sophisticated algorithms to mitigate modal dispersion.
Modal dispersion in MMF (even within the same mode group) can be 50$\times$ larger than polarization modal dispersion (PMD) in conventional SMF~\cite{MODALDISP}. Using dispersion as an alternative projection for phase retrieval does not rely on sparsity. It uses the same number of electrical signals as used in
coherent receivers  but eliminates the 90-degree optical hybrids and any CW carrier or LO, and replaces the balanced photo-diodes with simpler single-ended photo-diodes.

We demonstrate 360-Gbits/s (line rate) mode-multiplexed QPSK transmission over 30-km graded-index (GI) MMF supporting three spatial modes using direct detection.

Figure~\ref{FIG:1} shows the schematic of the receiver architecture for reconstructing the full field of six spatial and polarization modes: $LP_{01}$, $LP_{11a}$ and $LP_{11b}$, each with two polarizations.
After demultiplexing, each spatial and polarization mode is detected by two single-ended photo diodes.
The  two intensity measurements differ by the dispersion ($D$) added in one path.
To simplify the receiver, all the spatial and polarization modes can share one common component for the alternative projection, e,g., using a volume Bragg grating with a large aperture~\cite{Bragg}.


\begin{figure*}[!t]
	\vskip -18pt
	\centerline{\includegraphics[width=6.8in]{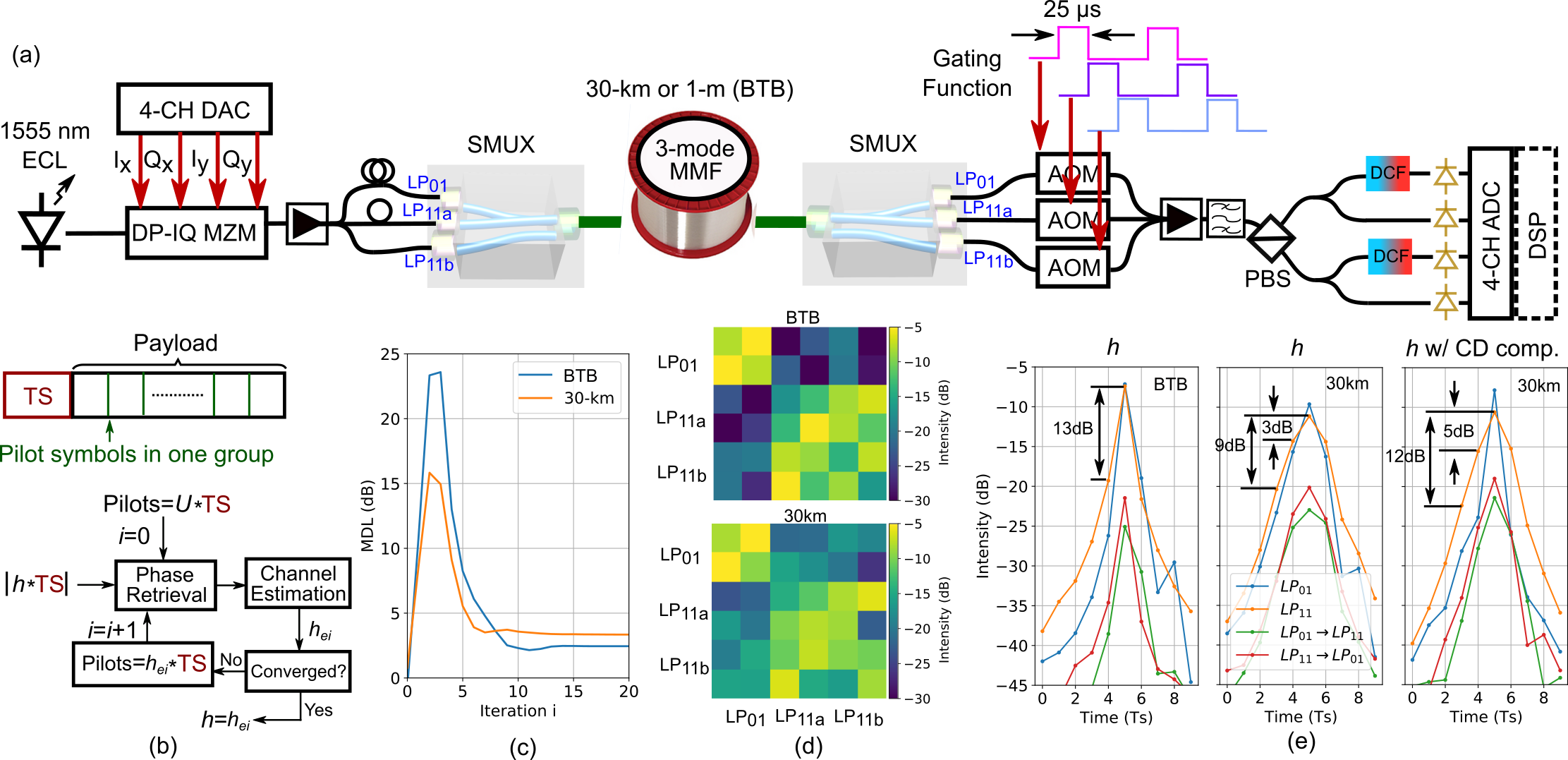}}
	\vspace{-2pt}
	\caption{(a)~Experimental setup for detecting the full field of the mode-multiplexed signals transmitted over the 3-mode MMF using only direct detection
		(b)~illustration of the transmitted signal frame structure and 6$\times$6 time-domain transfer matrix $h$ estimation (TS: training sequence, *: convolution, $U$: unitary matrix and $h_{ei}$: estimated $h$ at $i_{th}$ iteration),
		(c)~calculated MDL of $h_{ei}$,
		(d)~intensity of the 6$\times$6 estimated $h$,
		(e)~channel impulse responses of different mode groups for BTB and after 30-km transmission ($T_s$: symbol period).}
	\vspace{-8pt}
	\label{FIG:2}
\end{figure*}

\section{Experimental Setup}
Figure~\ref{FIG:2}(a) shows the experimental setup for detecting the full field of the mode-multiplexed signals transmitted over the 3-mode MMF and a time-multiplexed receiver to reduce the number of required ADC channels~\cite{TDM}.
An external cavity laser (ECL) with a linewidth of 100~kHz operating at 1555 nm is modulated with a  polarization-multiplexed 30-Gbaud QPSK signal with a spectral roll-off of 0.1 and pattern length of $2^{11}$, generated using a dual-polarization IQ Mach-Zehnder modulator (DP-IQ MZM).
It is driven by a 4-channel programmable digital-to-analog converter (DAC).
After amplification, three time decorrelated copies are sent to the single-mode side of a spatial multiplexer (SMUX) with a mode-group selectivity better than -15 dB~\cite{SMUX}.
We carried out experiments for both back-to-back (BTB) and 30-km transmission.
The 30-km 3-mode GI-MMF is a differential group delay (DGD) compensated span, which consists of two sections with opposite DGD.
After transmission, all the spatial modes are demultiplexed by second mode-group selective SMUX and gated by three acousto-optic modulators (AOM).
The gating function of the AOMs has a pulse width of 25 $\mu$s and 33.33$\%$ duty cycle. 
A continuous optical signal is created after recombining the three gated signals by applying time shifted gating functions.
The signal is captured by a phase retrieval receiver supporting two polarizations after being amplified and spectrally filtered.
Dispersion compensating fibers (DCFs) with dispersion of 650~ps/nm are placed in 2 of the four detector paths.
The four signals are direct detected and sampled by a 4-channel digital sampling oscilloscope operating at 60 GSamples/s. 
The transmitted signal frame structure for all the modes is illustrated in Fig.~\ref{FIG:2}(b).
The training sequence consists of 2048 symbols and is followed by a 0.5-million-symbol payload.

It takes two steps to perform phase retrieval over a MMF: 1) 6$\times$6 fiber transfer matrix \textit{$h$} estimation employing the training sequence and 2) full-field recovery over the payload, followed by the traditional digital signal processing (DSP) including 6$\times$6 MIMO.

\section{Transfer Matrix Reconstruction}
The 6$\times$6 MMF transfer matrix in time domain \textit{$h(n)$} is reconstructed through iteratively operating phase retrieval and channel estimation over the training sequence, see Fig.~\ref{FIG:2}(b).
For phase retrieval, the Gerchberg–Saxton (GS) algorithm is modified by adding a spectral constraint to force a rectangular spectrum for the Nyquist-shaped QPSK signal and pilot symbols to help with convergence.
The pilot symbols eliminate the constant phase ambiguity across the entire waveform, and assists the algorithm in converging to the optimum solution, especially in the presence of noise.
The channel estimation is implemented by multiplying the received training sequence with the pseudo inverse of time-aligned transmitted training sequence.
After each iteration, pilot symbols after MMF transmission are recalculated with $h_{ei}$, updated estimation of $h$.
$h$ can be properly estimated after 15 iterations.
Figure~\ref{FIG:2}(c) shows the calculated mode-dependent loss (MDL) as a function of the number of iterations.
The MDL is calculated based on the singular values of $h_{ei}$.
A unitary matrix ($U$) is used as the initial condition which produces a zero MDL at the $1^{st}$ iteration.
A slightly higher MDL is observed after 30-km transmission, mainly caused by the mode-profile mismatch between the MMF span and the pigtails of the SMUXes.
The intensity of the 6$\times$6 MMF transfer matrices are shown in Fig.~\ref{FIG:2}(d), where some mode coupling between the two mode groups can be observed after 30-km transmission.

The reconstructed $h$ contains both modal dispersion and CD: \textit{$h$} = \textit{$h_{CD}*h_{MD}$}, where \textit{$h_{CD}$} is treated as a common dispersion for all the modes and \textit{$h_{MD}$} is defined as the dispersion difference between the modes.
Figure~\ref{FIG:2}(e) provides the channel impulse responses within and between each mode group.
The time scale is normalized to the symbol period $T_s$.
Inter-symbol interference (ISI) is negligible in the BTB case, but becomes noticeable after 30-km transmission, caused by both CD and modal dispersion.
After applying CD compensation (see the right figure in Fig.~\ref{FIG:2}(e)), ISI is reduced.
However, the modal dispersion within the $LP_{11}$ mode group is still present and causes ISI.
To cope with the residual modal dispersion, after CD compensation, only \textit{$h_{MD}$} is used to forward-propagate the $M$ consecutive pilot symbols as one group, \textit{$\vec{p}$}, to the receiver where they are applied to correct the payload.
$M$=1 and 3 is set for BTB and 30-km transmission, respectively.
We have observed in both simulation and experiment that the initial dispersion \textit{$h_{CD}$} provides some additional amplitude randomness which offers more information to the phase retrieval algorithm and is beneficial in improving the convergence.
The spacing between pilot symbol groups can be reduced for recovering the signal after a transmission distance longer than 25 km.

\section{Transmission Results}
\begin{figure}[!t]
	\centerline{\includegraphics[width=2.5in]{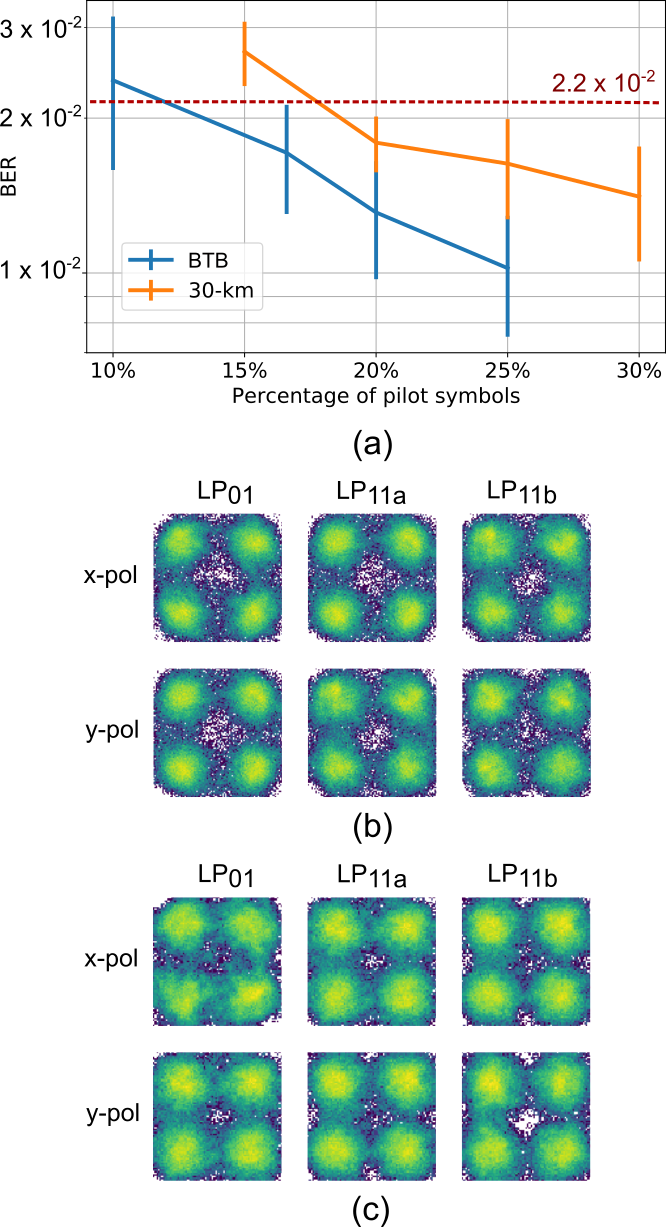}}
	\vspace{-4pt}
	\caption{(a) Achieved bit-error rate (BER) as a function of the percentage of the pilot symbols applied for phase retrieval,
		recovered QPSK constellations of all the six spatial and polarization modes employing 20$\%$ pilot symbols for (b) BTB and (c) 30-km GI-MMF transmission.}
	\vspace{-8pt}
	\label{FIG:3}
\end{figure}

Figure~\ref{FIG:3}(a) provides the mean and the variance of the calculated bit-error rate (BER) of all the modes.
The field of each spatial and polarization mode is first reconstructed using the block-wise phase retrieval algorithm, where a local-minima escape is operated after every 100 iterations~\cite{PR2}.
Parallel phase retrieval with different initial conditions helps to increase the convergence speed.
CD compensation, dispersive element forward and backward propagation and spectral filtering are all accomplished in the frequency domain using fast Fourier transform and its inverse.
After reconstructing the full fields of the mode-multiplexed signals, a 6$\times$6 frequency-domain MIMO equalizer with symbol-spaced taps is applied to recover the signals.
Figure~\ref{FIG:3}(b) and (c) shows the recovered QPSK constellations for all six spatial and polarization modes employing 20$\%$ pilot symbols for BTB and 30-km transmission, respectively.
Note that the system performance and the amount of pilot symbols applied is mainly limited by the waveform distortions due to the electrical noise introduced by the oscilloscope (ENOB) and lower than optimum photo-currents due to losses of the optical splitters and dispersive elements used in the time-multiplexed receiver.


\section{Conclusion and Outlook}
We presented the first carrier-less full-field recovery demonstration for mode-multiplexed signals (six spatial and polarization modes) using direct detection that shows the potential to use phase retrieval as an alternative receiver scheme for space-division multiplexing.
Compared to single-mode fiber transmission, performing phase retrieval over MMF is an order magnitude more complex because not only are the modes mixed over a larger dimensional space, but they also experience larger frequency-dependent variations due to modal dispersion. 
Moreover, mode-multiplexed signals are distorted due to MDL caused by the imperfections of the mode multiplexers and mode-profile mismatch.
We expect that the demonstrated receiver and phase retrieval scheme can be improved exploring parallel alternative projections~\cite{NEXT1}, integrating the algorithm with convex optimization~\cite{NEXT2} or machine learning~\cite{NEXT3}, and further optimizing the interaction between the phase-retrieval front-end and the traditional DSP blocks.

\end{document}